\documentclass{elsart}
\usepackage{amsfonts}

\def\1{{_{1}}}
\def\2{{_{2}}}
\def\3{{_{3}}}
\def\5{{_{5}}}
\def\7{{_{7}}}

\def\q{{q}}

\begin{document}

\runauthor{Jizba and Arimitsu}
\begin{frontmatter}
\title{Generalized statistics: yet another generalization
}
\author[Tsukuba]{Petr Jizba\thanksref{*}} and
\author[Tsukuba]{Toshihico Arimitsu}
\address[Tsukuba]{Institute of Physics, University of Tsukuba, Ibaraki
305-8571, Japan\thanksref{**}}
\thanks[*]{Corresponding author. Fax: +81--298--534492}
\thanks[**]{{\footnotesize{{\em E--mail addresses:} petr@cm.ph.tsukuba.ac.jp (P.J.),
arimitsu@cm.ph.tsukuba.ac.jp (T.A.).}} }

\begin{abstract}
\begin{minipage}{15.5cm}
We provide a unifying axiomatics for R\'{e}nyi's entropy and
non--extensive entropy of Tsallis. It is shown that the resulting
entropy coincides with Csisz\'{a}r's measure of directed
divergence known from communication theory.\\

\end{minipage}
\end{abstract}

\end{frontmatter}

\section{Introduction}\label{I}
It has been known already since Shannon's seminal
paper~\cite{Shannon1} that Shannon's information measure (or
enropy) represents mere idealized information appearing only in
situations when the buffer memory (or storage capacity) of a
transmitting channel is infinite. As the latter is not satisfied
in many practical situations, information theorists have invented
various remedies to deal with such cases. This usually consists of
substituting Shannon's information measure with information
measures of other types. Particularly distinct among them is a
one--parametric class of information measures discovered by
A.~R\'{e}nyi. It was later on realized by Linnik that these, so
called, R\'{e}nyi entropies (RE's) are associated to the decoding
limit if the source is compressed to ${\mathcal{I}}_q$ and the
parameter $q$ essentially tells how much the tail of a probability
distribution should count in the calculation of the R\'{e}nyi
entropy. Recently an operational characterization of RE in terms
of $\beta$--cutoff rates was provided by Csisz{\'a}r~\cite{Csi1}.

On the other hand, pioneering works of E.~Jaynes~\cite{Jaynes57}
in mid 50's revealed that the Gibbs entropy of statistical physics
represents the Shannon entropy whenever the sample space of
Shannon's entropy is identified with the set of all
(coarse--grained) microstates. However, contrary to information
theory, tendencies trying to extend the concept of Gibbs's entropy
have started to penetrate into statistical physics just recently.
This happened be realizing that there are indeed many situations
of practical interest requiring more ``exotic" statistics which do
not conform with the classical Gibbsian MaxEnt. Percolation,
polymers, protein folding, critical phenomena, cosmic rays,
turbulence or stock market returns provide examples.

One obvious way of generalizing Gibbs's entropy would be to look
on the axiomatic rules determining Shannon's information measure.
In fact, the usual axiomatics of Khinchin~\cite{Kh1} offers
various ``plausible" generalizations. The additivity of
independent mean information is then natural axiom to attack.
Along those lines only two distinct generalization schemes have
been explored in the literature so far. First consists of a
redefinition of the statistical mean and second generalizes
additivity rule. Respective entropies are then RE's~\cite{Re1} and
various deformed entropies~\cite{Naudts}. While RE's are natural
tools in statistical systems with a non--standard scaling
behavior, deformed entropies seem to be relevant to systems with
embedded non--locality. A suitable merger of the above
generalizations could provide a new conceptual playground suitable
for a statistical description of systems possessing both
self--similarity and non--locality. Examples being the early
universe cosmological phase transitions or currently much studied
quantum phase transitions. In this paper we attempt to merge  RE's
and Tsallis entropies.

\section{R\'{e}nyi's entropy: entropy of self--similar systems}
\label{IV}

As already mentioned, RE represents a step towards more realistic
situations encountered in communication theory. Among a myriad of
information measures RE's discern themselves by a firm operational
characterization given in terms of block coding and hypotheses
testing. R\'{e}nyi parameter $q$ then represents the so--called
$\beta$--cutoff rates~\cite{Csi1}. RE of order $q$ ($\q >0$) of a
discrete distribution ${\mathcal{P}} = \{ p_1, \ldots, p_n\} $
reads
\begin{eqnarray}
{\mathcal{I}}_q({\mathcal{P}}) = \frac{1}{(1 - q)} \ln\left(
\sum_{k=1}^n (p_k)^q  \right)\, .
\end{eqnarray}
Apart from coding theory RE's have proved to be indispensable
tools in various branches of physics. Examples being chaotic
dynamical systems or multifractals.
In his original work R\'{e}nyi~\cite{Re1} introduced a
one--parameter family of information measures ($=$RE)  which he
based on axiomatic considerations. In the course of time his
axioms have been sharpened by Dar\'{o}tzy~\cite{Dar1} and
others~\cite{Oth2}. Most recently it was proved in Ref.~\cite{PJ1}
that RE can be conveniently characterized by the following set of
axioms:
\begin{enumerate}
\item For a given integer $n$ and given ${\mathcal{P}} = \{ p_1,
p_2, \ldots , p_n\}$ ($p_k \geq 0, \sum_k^n p_k =1$),
${\mathcal{I}}({\mathcal{P}})$ is a continuous with respect to all
its arguments.

\item For a given integer $n$, ${\mathcal{I}}(p_1, p_2, \ldots ,
p_n)$ takes its largest value for $p_k = 1/n$ ($k=1,2, \ldots, n$)
with the normalization ${\mathcal{I}}\left( \frac{1}{2},
\frac{1}{2}\right) =\ln 2$.

\item For a given $q\in {{\mathbb{R}}}$ define $\varrho_k(q) =
(p_k)^{q}/\sum_k (p_k)^{q}$ (${\mathcal{P}}$ is affiliated to $A$)
then \\
${\mathcal{I}}(A\cap B) = {\mathcal{I}}(A) + {\mathcal{I}}(B|A)\ $
where $\ {\mathcal{I}}(B|A) = \mbox{{\textsl{g}}}^{-1}
\left(\sum_k \varrho_k(q)
\mbox{{\textsl{g}}}({\mathcal{I}}(B|A=A_k)) \right)$.

%
\item $\mbox{{\textsl{g}}}$ is invertible and positive in $[0,
\infty)$.

\item ${\mathcal{I}}(p_1,p_2, \ldots , p_n, 0 ) =
{\mathcal{I}}(p_1,p_2, \ldots , p_n)$.
\end{enumerate}

\noindent Former axioms markedly differ from those utilized
in~\cite{Re1,Dar1,Oth2}. One particularly distinct point is the
appearance of the escort distribution $\varrho(q)$ in axiom 3.
Note also that RE of two independent experiments is additive. In
fact, it was shown in Ref.~\cite{Re1} that RE is the most general
information measure compatible with additivity of independent
information and Kolmogorov axioms of probability theory.
%
%
\section{Tsallis entropy: entropy of long distance correlated
systems}\label{V}

Among variety of deformed entropies the currently popular one is
the $q$--additivity prescription and related Tsallis entropy (TE).
As the classical additivity of independent information is
destroyed in this case, a new more exotic physical mechanisms must
be sought to comply with TE predictions. One may guess that the
typical playground for TE should be cases when two statistically
independent systems have non--vanishing long--range/time
correlations: e.g., statistical systems with quantum
non--locality. In the case of discrete distributions
${\mathcal{P}} = \{ p_1, \ldots, p_n \}$ TE takes the form:
\begin{eqnarray}
{\mathcal{S}}_q({\mathcal{P}}) = \frac{1}{(1-q)} \left[
\sum_{k=1}^n (p_k)^q  -1\right]\, .
\end{eqnarray}
\noindent Axiomatic treatment was recently proposed in
Ref.~\cite{Ab2} and it consists of four axioms
\begin{enumerate}
\item For a given integer $n$ and given ${\mathcal{P}} = \{ p_1,
p_2, \ldots , p_n\}$ ($p_k \geq 0, \sum_k^n p_k =1$),
${\mathcal{S}}({\mathcal{P}})$ is a continuous with respect to all
its arguments.

\item For a given integer $n$, ${\mathcal{S}}({\mathcal{P}})$
takes its largest value for $p_k = 1/n$ ($k=1,2, \ldots, n$).

\item For a given $q\in {\mathbb{R}}$; ${\mathcal{S}}(A\cap B) =
{\mathcal{S}}(A) + {\mathcal{S}}(B|A) +
(1-q){\mathcal{S}}(A){\mathcal{S}}(B|A)$ with
 ${\mathcal{S}}(B|A) =
\sum_k \varrho_k(q) \ {\mathcal{S}}(B|A=A_k)$.

%

\item ${\mathcal{S}}(p_1,p_2, \ldots , p_n, 0 ) =
{\mathcal{S}}(p_1,p_2, \ldots , p_n)$.
\end{enumerate}
As said before, one keeps here the linear mean but generalizes the
additivity law. In fact, the additivity law in axiom~3 is nothing
but the Jackson sum
of the $q$--calculus.

%

\section{J--A axioms and solutions}\label{VII}

Let us combine the previous two axiomatics in the following
natural way:
\begin{enumerate}
\item For a given integer $n$ and given ${\mathcal{P}} = \{ p_1,
p_2, \ldots , p_n\}$ ($p_k \geq 0, \sum_k^n p_k =1$),
${\mathcal{D}}({\mathcal{P}})$ is a continuous with respect to all
its arguments.

\item For a given integer $n$, ${\mathcal{D}}({\mathcal{P}})$
takes its largest value for $p_k = 1/n$ ($k=1,2, \ldots, n$).

\item For a given $q\in {\mathbb{R}}$; ${\mathcal{D}}(A\cap B) =
{\mathcal{D}}(A) + {\mathcal{D}}(B|A) +
(1-q){\mathcal{D}}(A){\mathcal{D}}(B|A)$ with ${\mathcal{D}}(B|A)
= f^{-1}\left(\sum_k \varrho_k(q) \
f\left({\mathcal{D}}(B|A=A_k)\right)\right)$.

%

\item $f$ is invertible and positive in $[0, \infty)$.

\item ${\mathcal{D}}(p_1,p_2, \ldots , p_n, 0 ) =
{\mathcal{D}}(p_1,p_2, \ldots , p_n)$.
\end{enumerate}
We will now show that the above axioms allow for only one class of
solutions which will be closely related to the cross--entropy
measures of Havrda and Charvat~\cite{HaCh}.
%
%

\section{Basic steps in the proof}\label{VIII}

Let us first denote ${\mathcal{D}}(1/n,1/n, \ldots, 1/n) =
{\mathcal{L}}(n)$. Axioms $2$ and $5$ then imply that
${\mathcal{L}}(n) = {\mathcal{D}}(1/n,\ldots, 1/n,0) \ \leq \
{\mathcal{D}}(1/{n+1}, \ldots, 1/{n+1}) = {\mathcal{L}}(n+1)$.
Consequently ${\mathcal{L}}$ is a non--decreasing function. To
determine the form of ${\mathcal{L}}(n)$ we will assume that
${\mathcal{A}}^{(1)}, \ldots, {\mathcal{A}}^{(m)}$ are independent
experiments each with $r$ equally probable outcomes:
\begin{eqnarray}
{\mathcal{D}}({\mathcal{A}}^{(k)}) = {\mathcal{D}}(1/r, \ldots,
1/r) = {\mathcal{L}}(r)\, , \;\;\;\; (1 \leq k \leq m)\, .
\end{eqnarray}
Repeated application of axiom $3$ then leads to
\begin{eqnarray}
{\mathcal{D}}({\mathcal{A}}^{(1)}\cap {\mathcal{A}}^{(2)} \cap
\ldots \cap {\mathcal{A}}^{(m)}) \ &=& \ {\mathcal{L}}(r^m) \ = \
\sum_{i=1}^m  {m\choose i} (1- q)^{i-1}
{\mathcal{D}}^i({\mathcal{A}}^{(i)})\nonumber \\
&=& \ \frac{1}{(1- q)} \left[ \left( 1 + (1 - q){\mathcal{L}}(r)
\right)^m - 1\right]\, . \label{VI1}
\end{eqnarray}
Taking partial derivative of both sides of (\ref{VI1}) with
respect to $m$ and putting $m=1$ afterwards we get the
differential equation
\begin{eqnarray}
\frac{(1-q) \ d {\mathcal{L}}}{(1 + (1-q) \
{\mathcal{L}})\left[\ln\left(1 + (1-q)\ {\mathcal{L}}
\right)\right] } \ = \ \frac{dr}{r \ln r}\, . \label{VI2}
\end{eqnarray}
It is easy to verify that the general solution of (\ref{VI2}) has
the form
\begin{eqnarray}
{\mathcal{L}} (r) \ \equiv \ {\mathcal{L}}_q (r) \ = \
\frac{1}{1-q}\left( r^{c(q)} -1\right)
\, . \label{VIa}
\end{eqnarray}
Function $c(q)$ will be determined later on. Right now we just
note that because at $q = 1$ Eq.(\ref{VI1}) boils down to
${\mathcal{L}}(r^m) = m{\mathcal{L}}(r)$ we have $c(1) = 0$. We
proceed by considering the experiment with outcomes ${\mathcal{A}}
= ({\mathcal{A}}_1,{\mathcal{A}}_2, \ldots, {\mathcal{A}}_n)$ and
the distribution ${\mathcal{P}} = (p_1, p_2, \ldots, p_n)$. Assume
moreover that $p_k \, (1\leq k \leq n)$ are rational numbers,
i.e., $p_k = g_k /g$, $\sum_{k=1}^n g_k = g$  with $g_k \in
\mathbb{N}$. Let us have, furthermore, an experiment
${\mathcal{B}} = ({\mathcal{B}}_1, {\mathcal{B}}_2,\ldots,
{\mathcal{B}}_g)$ with distribution ${\mathcal{Q}} = \{q_1, q_2,
\ldots, q_g \}$. We split
$({\mathcal{B}}_1,{\mathcal{B}}_2,\ldots, {\mathcal{B}}_g)$ into
$n$ groups containing $g_1, g_2, \ldots, g_n$ outcomes
respectively. Consider now a particular situation in which
whenever event ${\mathcal{A}}_k$ happens then in ${\mathcal{B}}$
all  $g_k$ events of $k$-th group occur with the equal probability
$1/g_k$ and all the other events in ${\mathcal{B}}$ have
probability zero. Hence ${\mathcal{D}}({\mathcal{B}}|{\mathcal{A}}
= {\mathcal{A}}_k) = {\mathcal{D}}(1/g_k, \ldots, 1/g_k) =
{\mathcal{L}}_q( g_k)$ and so by axiom $3$ we have
\begin{eqnarray}
{\mathcal{D}}({\mathcal{B}}|{\mathcal{A}}) = f^{-1}\left(
\sum_{k=1}^n \varrho_k(q) f({\mathcal{L}}_q (g_k)) \right) \, .
\end{eqnarray}
\noindent On the other hand, for our system  the entropy
${\mathcal{D}}({\mathcal{A}}\cap {\mathcal{B}})$ can be easily
evaluated. Realizing that the joint probability distribution
corresponding to ${\mathcal{A}}\cap {\mathcal{B}}$ is
\begin{eqnarray*}
{\mathcal{R}} = \left\{ r_{kl} = p_k q_{l|k}
 \right\} = \{ \underbrace{\frac{p_1}{g_1}, \ldots,  \frac{p_1}{g_1}, }_{g_1
\times} \underbrace{\frac{p_2}{g_2}, \ldots,\frac{p_2}{g_2},
}_{g_2 \times} \ldots, \underbrace{\frac{p_n}{g_n}, \ldots,
\frac{p_n}{g_n} }_{g_n \times}\} = \left\{ 1/g, \ldots, 1/g
\right\}\, ,
\end{eqnarray*}
\noindent we obtain that ${\mathcal{D}}({\mathcal{A}}\cap
{\mathcal{B}}) = {\mathcal{L}}_q(g)$. Utilizing the first part of
axiom 3 and defining
%
%
$f_{(a,y)}(x) = f(-ax + y)$
we can write
\begin{eqnarray*}
{\mathcal{D}}({\mathcal{A}}) = \frac{f^{-1}_{(a,
{\mathcal{L}}(g))} \left( \sum_k \varrho_k(q) f_{(a,
{\mathcal{L}}(g))}(-{\mathcal{L}}_q (p_k)) \right)}{1-(1-q)
f^{-1}_{(a, {\mathcal{L}}(g))} \left( \sum_k \varrho_k(q) f_{(a,
{\mathcal{L}}(g))}(-{\mathcal{L}}_q (p_k)) \right)}\, , \;\;\; a =
[1 + (1-q) {\mathcal{L}}_q(g)]\, .
\end{eqnarray*}
%

As Eq.(\ref{VIa}) indicates it is ${\mathcal{L}}_q(1/p_k)$ and not
$-{\mathcal{L}}_q(p_k)$ which represents the elementary
information of order $q$ affiliated with $p_k$. Thus using the
relation
\begin{eqnarray}
{\mathcal{L}}_q(p_k) =
-\frac{\mathcal{L}_q(1/p_k)}{1+(1-q){\mathcal{L}}_q(1/p_k)}\, ,
\end{eqnarray}
together with transformation
\begin{eqnarray}
{\textsl{g}}(x) = f_{(a, {\mathcal{L}}(g))}\left(
\frac{x}{1+(1-q)x} \right)\, ,
\end{eqnarray}
we easily obtain that
\begin{eqnarray}
{\mathcal{D}}({\mathcal{A}}) = {\textsl{g}}^{-1}\left( \sum_k
\varrho_k(q) {\textsl{g}}({\mathcal{L}}_q \left(1/p_k\right))
\right) \ = \ f^{-1}\left( \sum_k \varrho_k(q) f({\mathcal{L}}_q
\left(1/p_k\right)) \right) \, . \label{VIb}
\end{eqnarray}
The last identity is due to second part of axiom $3$. It is well
known from the theory of means~\cite{HLP1} that (\ref{VIb}) can be
fulfilled iff ${\textsl{g}}(x) $ is a linear function of $f(x)$,
i.e.,
\begin{eqnarray}
{\textsl{g}}(x) = f\left( \frac{-x + y}{1+(1-q)x}\right) =
\theta_q(y) f(x) \ + \ \vartheta_{q}(y)\, . \label{VIc}
\end{eqnarray}
In order to solve (\ref{VIc}) we define $\varphi(x) = f(x) -f(0)$.
With this notation (\ref{VIc}) turns into
\begin{eqnarray}
\varphi\left(\frac{-x + y}{1+(1-q)x}\right) = \theta_q(y)
\varphi(x) \ + \ \varphi(y)\, , \; \; \;\; \; \varphi(0) = 0\, .
\label{VId}
\end{eqnarray}
By setting $x$ = $y$  we see that $\theta_q(y) = -1$, hence one
finds
\begin{eqnarray}
\varphi(x + y + (1-q)xy) \ = \ \varphi(x) + \varphi(y)\, .
\label{VI3}
\end{eqnarray}
Eq.(\ref{VI3}) is Pixeder's functional equation which can be
solved by the standard method of iterations~\cite{Acz1}.
Eq.(\ref{VI3}) has only one non--trivial class of solutions,
namely:
\begin{eqnarray}
\varphi(x) = \frac{1}{1-\alpha} \ \ln\left(  1+ (1-q)x \right)\, .
\end{eqnarray}
$\alpha$ is here a free parameter. Plugging this solution back to
(\ref{VIb}) we obtain
\begin{eqnarray}
{\mathcal{D}}_q({\mathcal{A}}) &=&   \frac{1}{1-q}\ \left(
e^{-c(q)\sum_k \varrho_k(q) \ln p_k}  -1 \right) = \frac{1}{1-q}
\left( \prod_k (p_k)^{-c(q)\varrho_k(q)} -1 \right)\, .
\label{VIe}
\end{eqnarray}
Note that the constant $\alpha$ got cancelled. It remains to
determine $c(q)$. Utilizing the conditional entropy constructed
from (\ref{VIe}) and using axiom $3$, we obtain $c(q) = 1-q$.
Inasmuch we can recast (\ref{VIe}) into more expedient form
\begin{eqnarray}
{\mathcal{D}}_q({\mathcal{A}}) =  \frac{1}{1-q}\ \left(
e^{-(1-q)^2 d{\mathcal{I}}_q/dq} \sum_{k=1}^n (p_k)^q -1 \right)\,
. \label{VIf}
\end{eqnarray}
Eqs.(\ref{VIe})--(\ref{VIf}) are the sought results. In passing we
note that ${\mathcal{D}}_q \geq 0$ for $\forall q \in \mathbb{R}$
and $\lim_{q\rightarrow1}{\mathcal{D}}_q  = {\mathcal{I}}_1 =
{\mathcal{S}}_1$.

\vspace{-5mm}
\section{Conclusions and outlooks}\label{IX}
\vspace{-5mm}

Presented axiomatics might provide a novel playground for a $q$
non--extensive systems with embedded self--similarity. Indeed, one
could expect that the obtained measure of information could play a
relevant r\^{o}le in $q$ non--extensive statistical systems near
critical points. Research in this direction
is currently in progress.

A curious result arises when one restricts values of
${\mathcal{P}}$ by the constraint
$d{\mathcal{I}}_q({\mathcal{P}})/dq = \max_{p_i}
{\mathcal{I}}_q({\mathcal{P}})/(1-q)$. Eq.(\ref{VIf}) then boils
down to
\begin{eqnarray}
{\mathcal{D}}_q({\mathcal{A}}) =  \frac{1}{1-q}\ \left( n^{q-1}
\sum_{k=1}^n (p_k)^q -1 \right) \equiv -
{\mathcal{C}}_q({\mathcal{A}})\, . \label{VIg}
\end{eqnarray}
The reader may recognize in ${\mathcal{C}}_q$ the generalized
measure of cross--entropy of Havrda and Charvat~\cite{HaCh} (also
known as Csisz\'{a}r's measure of directed divergence~\cite{Csi2})
used in communication theory. For $q=2$ we recover the $\chi^2$
measure. This suggests that the non--extensivity together with
self--similarity may be important concepts also in information
theory. In this connection such issues as the channel capacitance
and cutoff rates would deserve a separate discussion.

The generalized entropy ${\mathcal{D}}_q$ has many desirable
features: like Tsallis entropy it satisfies the non--extensive
$q$--additivity, involves a single parameter $q$, and goes over
into the standard Shannon entropy in the limit $q \rightarrow 1$.
On that basis it would appear that both ${\mathcal{S}}_q$ and
${\mathcal{D}}_q$ have an equal right to furnish a generalization
of statistical mechanics.




\end{document}